\documentclass[vecphys]{svmult}

\usepackage{makeidx}      
\usepackage{graphicx}
\usepackage{multicol}        
\usepackage[bottom]{footmisc}
\makeindex            
\newcommand{\farcm}{\mbox{\ensuremath{.\mkern-4mu^\prime}}}
\newcommand{\farcs}{\mbox{\ensuremath{.\!\!^{\prime\prime}}}}

\begin{document}

\title*{Deepest Near-IR Surface Photometry of Galaxies in the Local Sphere of Influence}
\titlerunning{Deep Near-IR Photometry of Local Volume Galaxies}
\author{Emma Kirby\inst{1}\and
Helmut Jerjen\inst{1}\and
Stuart Ryder\inst{2}\and
Simon Driver\inst{3}}
\authorrunning{Kirby et al.}

\institute{Research School of Astronomy and Astrophysics, Australian National University
\texttt{emma@mso.anu.edu.au, jerjen@mso.anu.edu.au}
\and Anglo-Australian Observatory \texttt{sdr@aao.gov.au}
\and School of Physics and Astronomy, University of St. Andrews \texttt{spd3@st-and.ac.uk}}

\maketitle
\begin{abstract}
We present near-IR, deep (4 mag deeper than 2MASS) imaging of 56 Local Volume galaxies. Global parameters such as total magnitudes and stellar masses have been derived and the new near-IR data combined with existing 21cm  and optical $B$-band data. We present multiwavelength relations such as the HI mass-to-light ratio and investigate the maximum total baryonic mass a galaxy can have.
\end{abstract}

\section{Photometry Beyond the 2MASS Limit}
\label{EKirby.sec1}

$H$-band ($1.6 \mu$m) images of 56 local galaxies ($D<10$ Mpc) were obtained with the Infrared Imager and Spectrograph 2 (IRIS2) at the 3.9m Anglo-Australian Telescope (AAT) between October 2004 and September 2006.  IRIS2 has a Rockwell HAWAII--1 HgCdTe array with a pixel scale of $0\farcs45$~pixel$^{-1}$ and a $7\farcm7 \times 7\farcm7$ field-of-view.  The total on-source integration time was 30 min with a mean seeing of $1\farcs 3$.

The data reduction was carried out using the ORAC-DR pipeline within the {\sc starlink} package. Instrumental magnitudes for field stars were obtained employing standard IRAF routines and were cross-correlated with the 2MASS Point Source Catalog  to provide photometric calibration. The measured total $H$-band luminosity of each galaxy was converted into a stellar mass by adopting a mass-to-light ratio of $\Upsilon_{\ast}^H = 1.0~{\mathcal{M}}_{\odot} / L_{\odot}$. A detailed description the observations and analysis can be found in Kirby et al~\cite{EKirby.kirby07}. 

\section{Sharing the Baryons}
The relationship between the stellar mass (from optical or near-IR measurements) and the gas (from 21cm line observations) in a galaxy provides insight into galaxy evolution. We use our data combined with data from the literature to examine the HI mass-to-light ratio as well as the relationship between the HI mass and the stellar mass. The literature data that we include is coming from Virgo Cluster galaxies listed in the Goldmine database~\cite{EKirby.gavazzi03}. We also include the $B$-band data of Warren et al.~\cite{EKirby.warren06} and Bouchard et al.~\cite{EKirby.bouchard07} which have been transformed to the $H$-band employing the optical - near infrared magnitude transformation discussed in Kirby et al.~\cite{EKirby.kirby07}. This allows us to investigate this relationship for a wide range of galaxy morphologies.
\begin{figure}
\centering
\begin{tabular}{cc}
\mbox{\includegraphics[height=5.8cm]{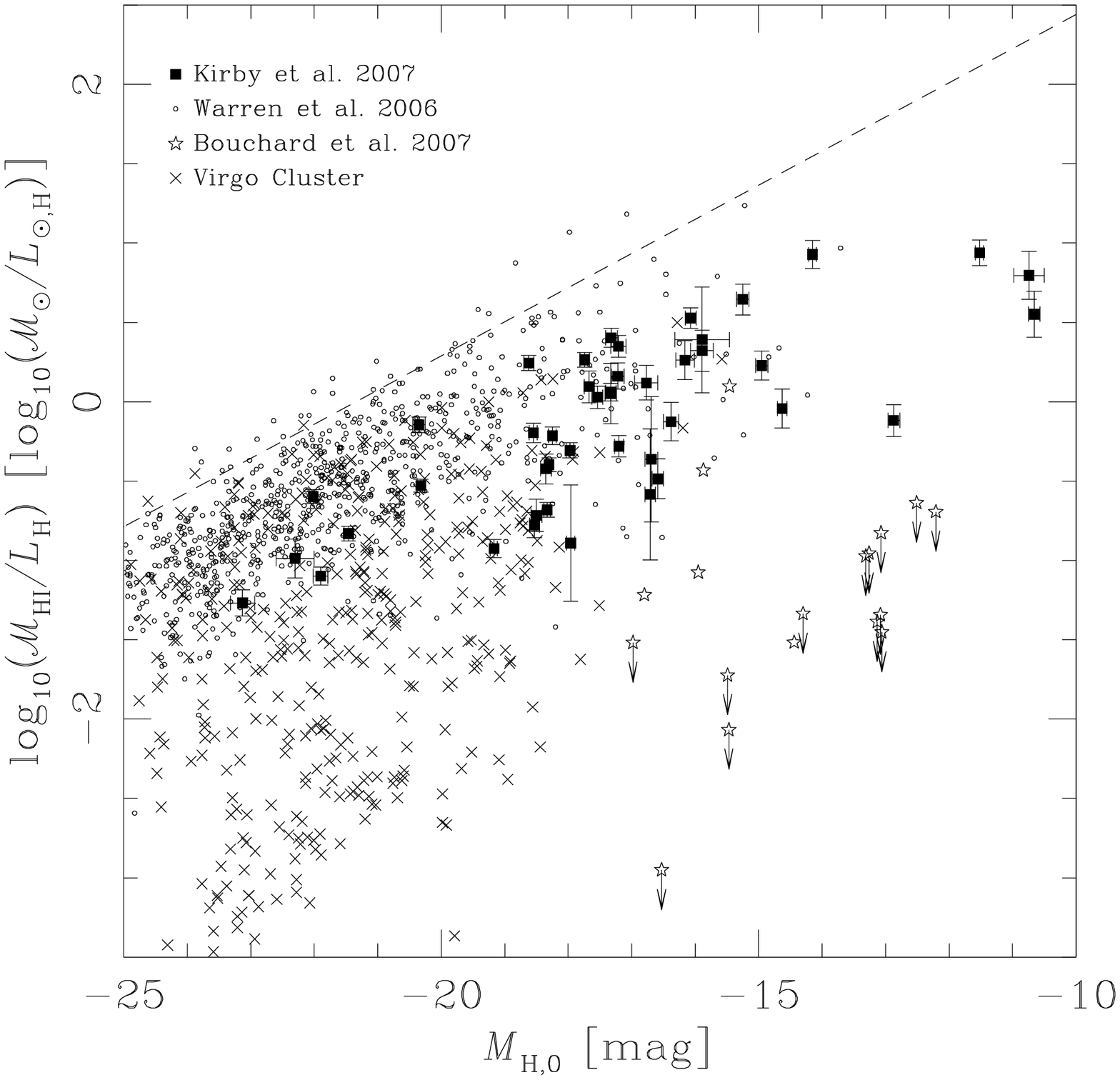}}&
\mbox{\includegraphics[height=5.8cm]{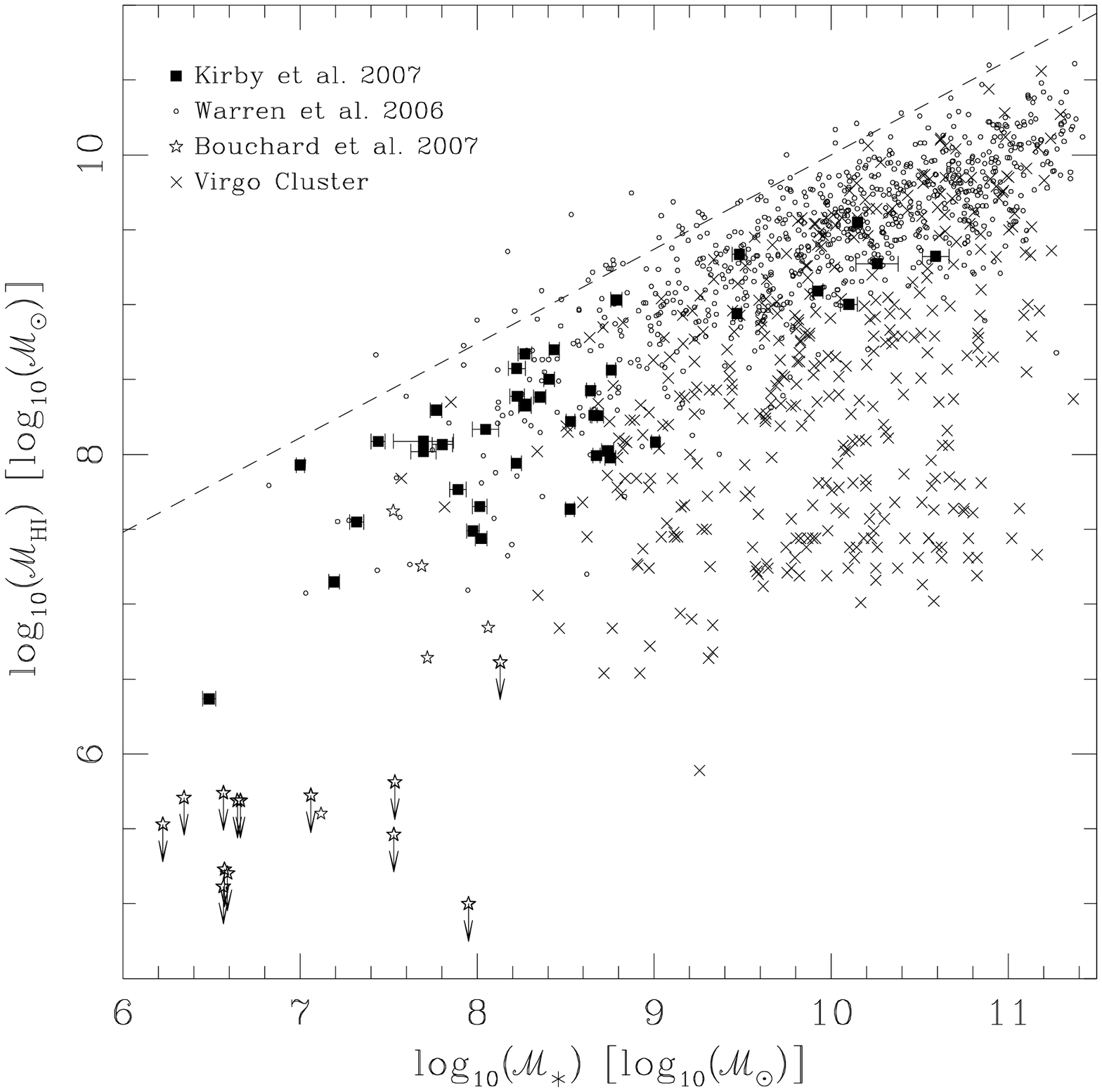}}\\
\end{tabular}
\caption{{\it{Left Panel}}: The HI mass-to-light ratio versus absolute $H$ band magnitude diagram. {\it{Right panel}}: The HI mass versus stellar mass plot. In both plots the dashed line represents the upper envelope to a galaxy's potential mass-to-light ratio.}
\label{EKirby.fig.M2L}
\end{figure}

\subsection{The HI mass-to-light ratio}
The HI mass-to-light ratio measures the relative amounts of gas and stars in a galaxy. In  Fig.~\ref{EKirby.fig.M2L} (left panel) we show the HI mass-to-light ratio for the selected galaxies.  Warren et al.~\cite{EKirby.warren06} proposed  that there is an upper envelope to a galaxy's mass-to-light ratio at a given luminosity (dashed line). This suggests that the minimum amount of stars that a galaxy will form is related to the initial baryonic mass, a hypothesis supported by the theoretical work of Taylor \& Webster~\cite{EKirby.taylor05}. For a galaxy to lie above this upper envelope, it would have to be very massive yet also have low surface brightness (to remain optically undetected) and low HI column density. However, HIDEEP~\cite{EKirby.minchin03} found no giant LSB galaxies and no galaxies with HI column density less than $10^{20}$\,cm${}^{-2}$ despite a sensitivity limit 1.5\,dex lower than this. 

\subsection{HI mass versus stellar mass}
Although the HI mass-to-light ratio is frequently used to describe the relationship between the gas and the stars in a galaxy, when examining Fig.~\ref{EKirby.fig.M2L} (left panel), it is important to note that the two parameters plotted are dependent on each other. A better way to look at this is presented in the right hand panel of Fig.~\ref{EKirby.fig.M2L} where we show how the HI mass relates directly to the stellar mass.

Our sample, containing late-type dwarfs, is closer to the upper envelope than the early-type dwarf sample of Bouchard et al.~\cite{EKirby.bouchard07}. Similarly, the sample of  field galaxies of Warren et al.~\cite{EKirby.warren06} is much closer to the upper envelope than the Virgo Cluster data that is dominated by Es and dEs. This highlights that the relationship between the stellar and gas masses is dependent on the environment and hence implicitly on the morphological type.

In the HI mass versus stellar mass plot, the upper envelope clearly does not have a gradient of unity. A 45${}^{\circ}$ line represents galaxies with exactly half their baryonic mass in stars and half in gas. High mass galaxies lie systematically below a 45${}^{\circ}$ line whereas low mass galaxies can fall either above or below a 45${}^{\circ}$ line. This means that the fraction of baryonic mass in stars is dependent on the total baryonic mass.

\subsection{Stellar mass fraction versus total baryonic mass}

\begin{figure}[!htb]
\centering
\includegraphics[height=8.8cm]{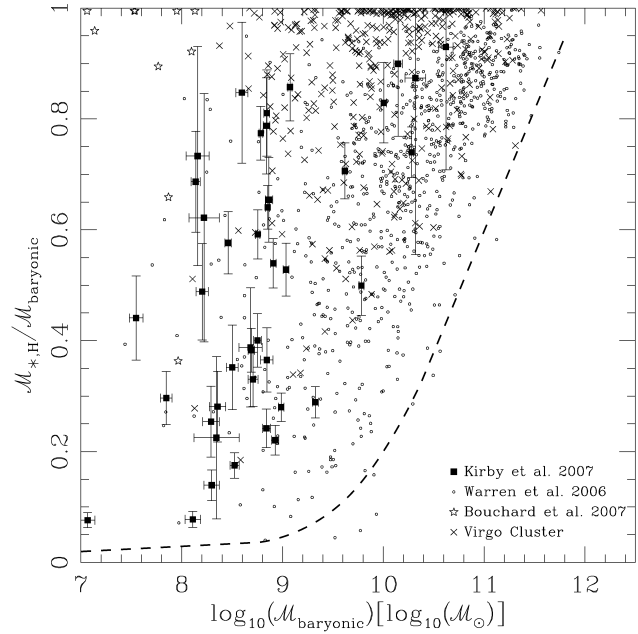}
\caption{Relation between the stellar mass fraction and the total baryonic mass in local galaxies. The dashed line is the empirical minimum mass fraction possible for a galaxy with a given total baryonic mass.}
\label{EKirby.fig.bary}
\end{figure}

In Fig.~\ref{EKirby.fig.bary}, a direct comparison between the stellar fraction of the baryonic content and the total baryonic mass  in a galaxy is given.
The empirically determined dashed line represents the minimum mass fraction required for the galactic disk to remain stable against star formation. It seems that at lower masses, galaxies can stabilise with less of their gas converted into stars. This is again supported by the theoretical work of Taylor and Webster~\cite{EKirby.taylor05}.

Fig.~\ref{EKirby.fig.bary} shows that the minimum stellar fraction of the baryonic mass goes to 1 as the total baryonic mass approaches $10^{12} {\mathcal{M}}_{\odot}$  suggesting that there is an upper limit to the total baryonic mass a galaxy can have. This upper limit is in good agreement with the observed upper limit of high redshift galaxies ($0.01<z<4$) found by Rocca-Volmerange et al.~\cite{EKirby.rocca04} and the theoretical value discussed by Rees and Ostriker~\cite{EKirby.rees77}.

\section{Conclusion}
Comparisons of the gas mass and the stellar mass provides insight into galaxy evolution. Galaxies with high baryonic mass-to-light ratios are often cited as solutions to ongoing cosmological problems, in particular reconciling the faint-end of the galaxy luminosity function with the predicted CDM Halo Mass Function. 

The maximum total baryonic mass of a galaxy (both observational and theoretical) corresponds to the mass of a galaxy which has converted all of its gas into stars. Our empirical results strongly suggest an upper baryonic mass limit of $10^{12} {\mathcal{M}}_{\odot}$.

\subsection*{Acknowledgements} 
The authors acknowledge financial support from the Australian Research Council via the grant DP0451426.

\printindex
\end{document}